# Preparation and physical properties of soft magnetic nickel-cobalt nanowires with modulated diameters

Sebastian Bochmann,[a] Dirk Döhler,[a] Beatrix Trapp,[b] Michal Stano,[b] Olivier Fruchart,[b,c] Julien Bachmann[a,d,*]

[a] *Friedrich-Alexander University of Erlangen-Nürnberg, Inorganic Chemistry, Egerlandstrasse 1, 91058 Erlangen, Germany*

[b] *Univ. Grenoble Alpes, CNRS, Grenoble INP, Inst NEEL, 38000 Grenoble, France*

[c] *Univ. Grenoble Alpes, CNRS, CEA, Grenoble INP, INAC-Spintec, 38000 Grenoble, France*

[d] *Institute of Chemistry, Saint Petersburg State University, 26 Universitetskii Prospect, Saint Petersburg, Petergof 198504, Russia*

**Abstract**

We establish a method to produce cylindrical magnetic nanowires displaying several segments, with a large versatility in terms of segment diameter and length. It is based on electroplating in alumina templates, the latter being prepared by several steps of anodization, wet etching and atomic layer deposition to produce, widen or shrink pores, respectively. We propose an analytical model to analyze the in-plane and out-of-plane magnetization loops of dense assemblies of multisegmented wires. The model considers inter-wires dipolar fields, end-domain curling and predicts the switching field of individual wires with no adjustable parameter. Its ingredients are crucial to extract reliable parameters from the fitting of loops, such as magnetization or the porosity of the array.

## Introduction

The continuously decreasing size of magnetic bits in hard disk drives has reached a dimension close to the fundamental superparamagnetic limit, which corresponds to the minimal volume of particles, or "grains", of a certain material needed to maintain a stable magnetic moment.[1] In parallel, magnetic random access memory (MRAM) is shifting from a concept to commercial devices combining the access speed and compactness of a solid-state device with the non-volatility of magnetic materials. Still, this remains a two-dimensional technology and as such faces technological and fundamental limitations. One possibility to overcome this ceiling and increase the areal storage density beyond 1 Tbit/in² is to turn to a three-dimensional storage platform. This is addressed by the concept of the racetrack memory by Parkin *et al.*, in which bits are stored as series of domain walls in dense arrays of tall, 'vertical' wires of magnetic material.[2,3] For the sake of simple and inexpensive implementation, there should be only one read/write cell per wire, located for example at the wafer surface. This requires that domain walls be set in motion along the wire, or "track", in order to bring the individual bits outside of the matrix. This could in principle be achieved by spin transfer torque created by the spin polarization of conduction electrons.[4–7] One major challenge in this respect is the controlled motion of magnetic domain walls in a digital manner to predefined positions along the wire. Otherwise, after a series of individual motions in a smooth, unstructured wire, the spread of wall mobility may cause domain walls to meet, interact, and possibly annihilate each other.[8] Creating pinning sites along the wire length is a strategy to digitize it by defining a finite number of well-identified domain wall positions. These sites can be experimentally realized by a modulation of the material composition and properties,[9] or by a modulation in the wire structure, such as the diameter.[8] In planar systems, 'notched' strips produced by standard lithographic methods have indeed exhibited successful pinning of domain walls in motion.[10] However, a transfer of this concept to a three-dimensional system is yet to be demonstrated.

Individual tools needed towards this goal are available from the literature. Straights wires are routinely prepared by electrodeposition of a magnetic metal in an appropriate template, such as 'anodic' alumina.[11] The synthesis of anodic alumina templates with diameter modulations has been demonstrated, as well, using periodic oscillations of the anodization voltage (or current density). However, those systems feature gradual diameter changes instead of abrupt ones (which is deleterious for domain wall pinning), and limited freedom for separating modulations along the pore length.[12–15] A similar system of 'anodic' pore arrays featuring a small number of well-defined, abrupt changes has been obtained by the combination of successive "mild" and "hard" anodization.[16] The limitation of this system is that the diameters of the individual segments cannot be tuned at will, since the interpore distance of mild and hard anodization must be adjusted to each other to avoid instabilities.

In this paper, we present a preparative method for generating parallel arrays of magnetic nanowires featuring abrupt changes in diameter at well-defined positions along their length and arbitrarily defined values of the diameters. This is achieved by the combination of several steps involving the anodization of aluminum (in mild conditions), isotropic pore widening, and atomic layer deposition (ALD). Subsequently, we perform electrodeposition of the modulated magnetic nanowires. The magnetization hysteresis of our samples, measured as ensembles of macroscopic size, displays a complex behavior related to the distribution of wire segment lengths, diameters, and dipolar interactions. We propose a model able to reproduce both the hysteretic and reversible parts of these loops, from which accurate geometric and magnetic material parameters can be extracted.

## Results and Discussion

### 1. Preparation of modulated wires

The anodization of aluminum enables the experimentalist to generate parallel arrays of straight pores perpendicular to the surface, with well-defined diameters that can be set to values between 20 nm and 400 nm and pore lengths between 0.5 µm and 100 µm.[17] Anodized aluminum has been abundantly used as a template for further elaboration of nanowires and nanotubes by galvanic and other methods.[18] While the ratio of pore diameter versus pitch cannot be varied much through anodization, the pore diameter can be adjusted after growth by an isotropic wet chemical etching step.[18,19] Conversely, atomic layer deposition (ALD) can be used to reduce the diameter of the template's pores after growth, or confer them with specific chemical or physical properties.[20–24] Atomic layer deposition is a thin film deposition technique from the gas phase, which is uniquely suited to coating three-dimensional structures, including deep pores with high aspect ratio.[25,26]

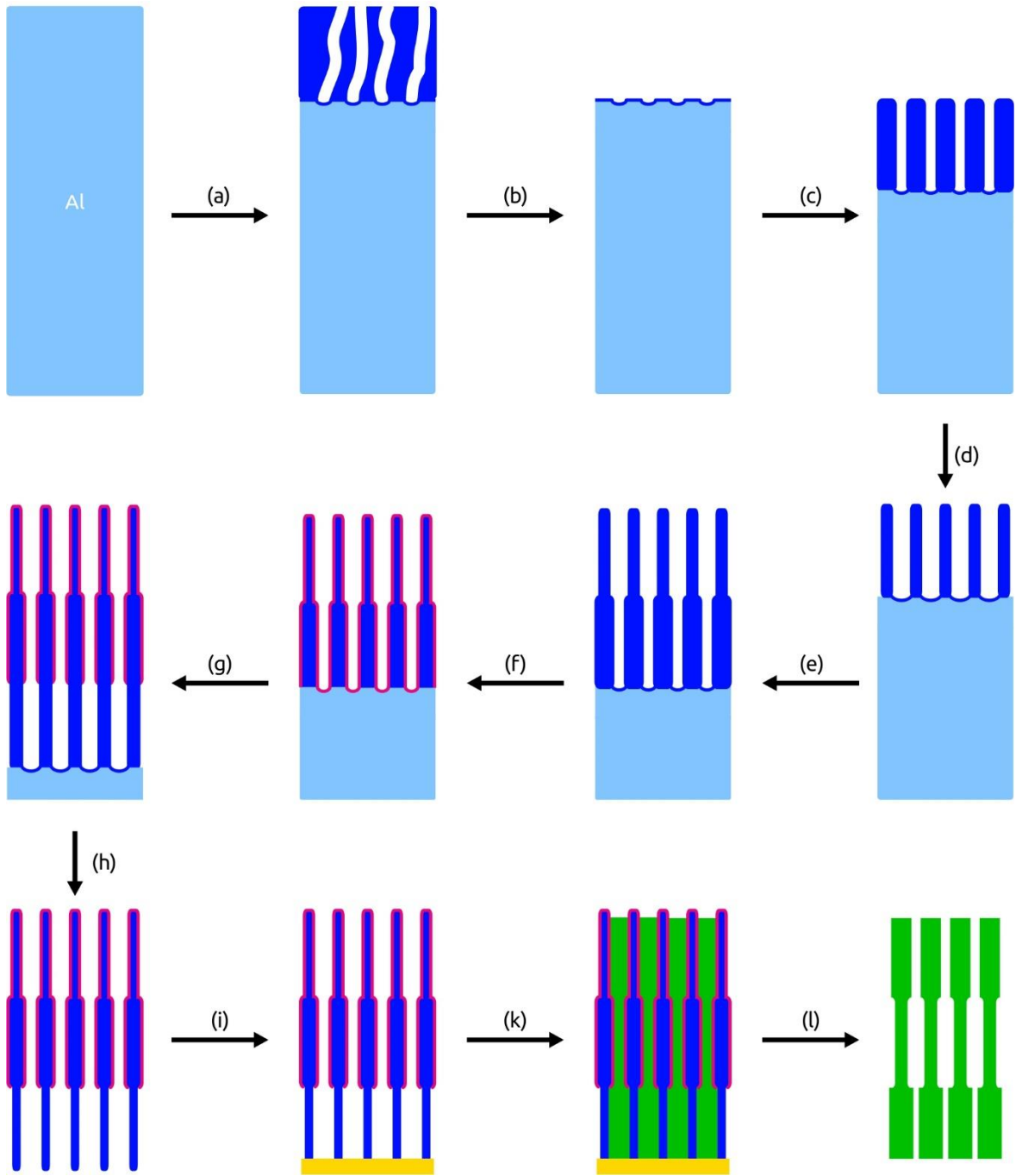

**Figure 1.** Sketch of the preparative procedure devised.

We combine these methods to generate modulated magnetic wires as sketched in **Figure 1**. After the first, sacrificial anodization step (a-b) used to generate the order, a first segment grown in 1% phosphoric acid electrolyte for 4 h exhibits a pore diameter of 150 nm and length of approximately 10 µm (c) according to standard procedures.[19] The pore is then submitted to isotropic widening in 10% phosphoric acid (d) to reach a diameter chosen here to be 400 nm, for example. A second anodization is performed under the same conditions (e), yielding a second segment with the natural diameter of 150 nm. The resulting pore structure featuring two distinct diameters is then coated with 5 nm $SiO_2$ via ALD (f).[27] If this coating is sufficiently thin, a third anodization can be performed to grow a third segment (g).[28] After removing the remaining metallic aluminum substrate and the $Al_2O_3$ barrier layer at the lower pore extremity (h), the diameter of the third segment can finally be increased by an isotropic etching step, for instance to match the first segment at 400 nm, whereas the first and second segments are protected by the chemically inert nature of the $SiO_2$ layer.

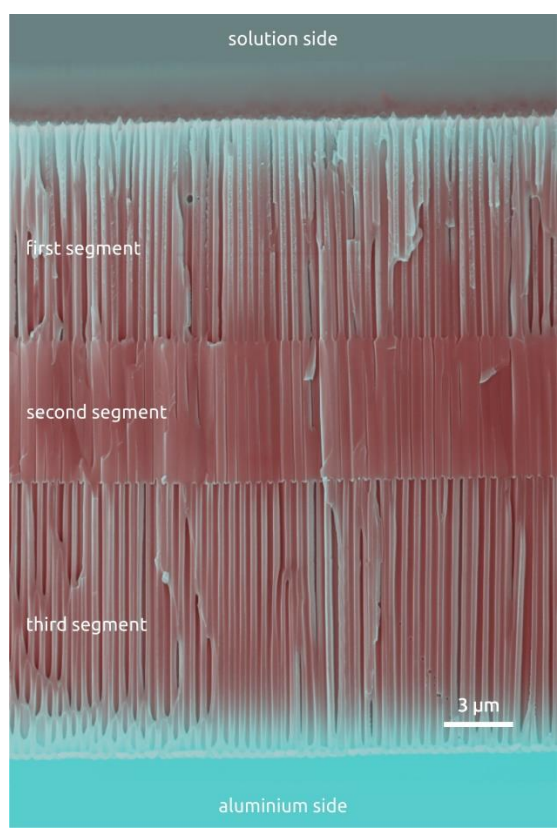

**Figure 2.** Scanning electron micrograph of a modulated template obtained with the method presented in **Figure 1**, with segment diameters of 400 nm, 150 nm and 400 nm. The false-color image is obtained from a combination of secondary and backscattered electron signals.

**Figure 2** shows the template structure obtained by this procedure. The pores are straight and parallel, the segments have a homogeneous length, and the transitions between distinct diameter values are abrupt. Note that our procedure is very flexible, in that each geometric parameter of the modulated structure (that is, each segment length and diameter) can be adjusted individually and independently of the others. For example, the diameter of the third segment can be either increased by pore widening, or reduced by $Al_2O_3$ ALD. The length of each segment can be adjusted by the anodization duration. If anodization is performed in 0.3 M oxalic acid (instead of 1% phosphoric acid), the initial pore diameter is 40 nm (instead of 150 nm) and can be widened to 90 nm (instead of 400 nm).[11]

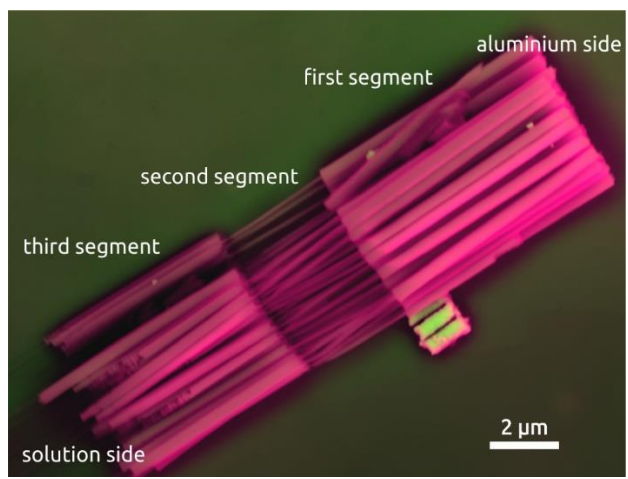

**Figure 3.** Scanning electron micrograph of a bundle of modulated NiCo wires ("nano-q-tips") isolated from the template, in false colors (combination of secondary and backscattered electron signals). NiCo appears pink, the polished Cu substrate dark green, and short Au stubs from the electrical contact neon green.

In the next step, an electrical contact is defined on one side of the alumina template by sputter-coating (i). We deposit 6 nm aluminum metal (as an adhesion layer) followed by 50 nm gold. This thin layer is then augmented by galvanic plating, performed in a two-electrode setup under –2.50 V in a $K[Au(CN)_2]$ electrolyte. The thick galvanic layer closes the pore extremities completely, and results in a short (1 μm long) Au stub inside the pores.[9,29] The sample is then exposed with the open pore extremity to an electrolytic solution containing cobalt and nickel. Soft magnetic $Ni_{60}Co_{40}$ alloy nanowires are then grown (k) in the modulated pores from the gold contact in a three-electrode configuration under –1.10 V vs. Ag/AgCl.[30] The resulting wires may optionally be removed from the template with chromic acid (l). **Figure 3** highlights how the diameter modulations of the template are reproduced in the galvanic wires with high fidelity. The characteristic shape suggests the nickname 'nano-q-tips' for such modulated wires. Backscattered electron contrast allows for differentiation between the long, modulated $Ni_{60}Co_{40}$ wires (pink) and a few broken off Au stubs (neon green).

As mentioned above, the diameter of each segment can be adjusted very precisely. **Figure 3** displays a rather extreme diameter "contrast" of 400/150/400 nm, but shorter durations of the isotropic pore widening steps allow one to generate more modest contrasts, for example the 200/150/200 nm sample presented in **Figure 4a**. Alternatively, performing the anodization steps in oxalic acid electrolyte instead of phosphoric acid enable one to work with smaller diameters. **Figure 4b** displays an excerpt of one such wire of the type 70/40/70 nm. This range of diameters is of interest in the magnetic realm since it is on the order of ten times typical magnetic exchange lengths, so that several topologies of domain walls may be expected.[31]

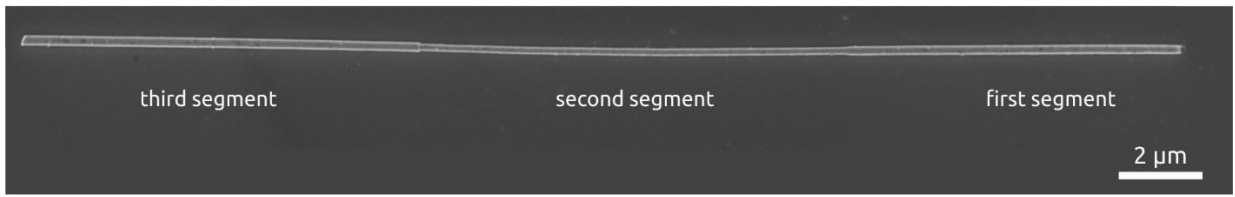

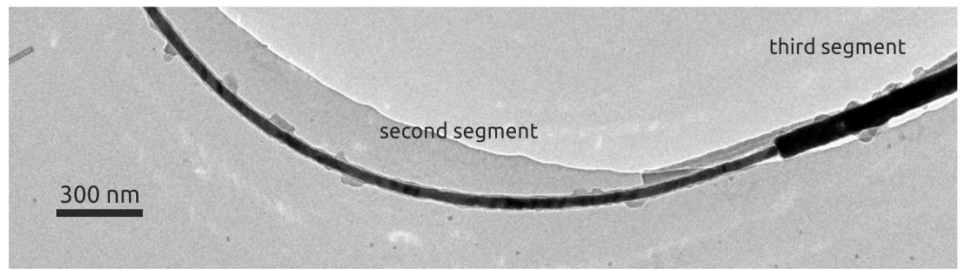

**Figure 4.** (a) Scanning electron micrograph of an isolated NiCo "nano-q-tip" with modest diameter contrast between segments of 200 nm, 150 nm, and 200 nm. (b) Transmission electron micrograph of a "nano-q-tip" isolated from pores grown in oxalic acid instead of phosphoric acid: The diameters are 70 nm, 40 nm, and 70 nm.

2. **Geometry of the diameter modulations**

The diameter reduction between first and second segment and the diameter increase between second and third segment result in distinct pore profiles (**Figure 5**). This is due to the distinct experimental procedures for diameter decrease and increase, as sketched in **Figure 5c** and **5d**. At the diameter <u>reduction</u>, the hemispherical extremity of a wide pore concentrates the electric field at the start of the subsequent anodization, so that the second segment continues straight from its lower extremity. This second segment immediately recovers its natural, narrower diameter, but does not affect the gradual, approximately hemispherical profile left at the end of the first segment. In contrast to this, the diameter <u>increase</u> occurs abruptly given that the hemispherical end of the narrow pore is lost completely upon growth of the next segment. The isotropic pore widening performed as the last preparative step finds an abrupt etch stop defined by the $SiO_2$ layer.

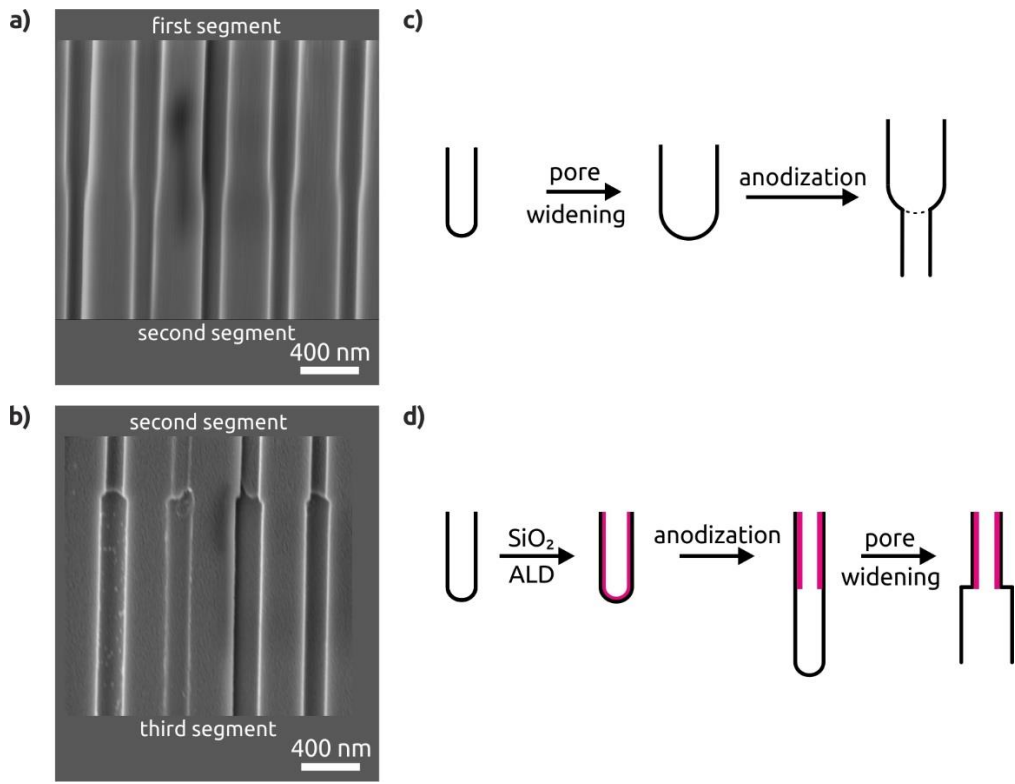

**Figure 5.** Transitions between two different pore diameters: Scanning electron micrographs highlighting the geometric difference between the gradual, conical diameter reduction from 200 nm to 150 nm (a) and the abrupt diameter increase from 150 nm to 200 nm (b). The corresponding anodization procedure for diameter decrease (c) and increase (d) is sketched.

Wires grown inside the modulated template exhibit the same geometric properties as the template (**Figure 6**). They exhibit diameter transitions with two distinct geometries: a smoother one obtained upon diameter reduction of the template pores stands in stark contrast to the abrupt one yielded by pores submitted to a diameter increase.

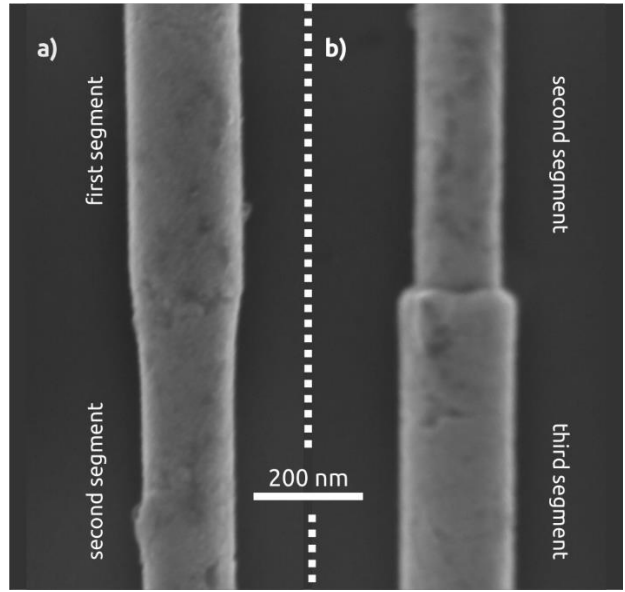

**Figure 6.** Transitions between two different wire diameters: (a) Gradual, conical diameter reduction; (b) Abrupt diameter increase.

3. **Experimental magnetic properties of modulated wires**

Magnetic nanowires display highly anisotropic magnetic properties, as the magnetization of soft magnetic materials tends to align along the long dimensions of an object.[32] When homogeneously magnetized wires are submitted to an external field oriented antiparallel to the magnetization, its reversal implies the formation of a magnetic domain wall, which nucleates at one extremity and then travels to the other. Its energy of formation, and thereby the reversal field measured experimentally, is strongly dependent on the wire diameter.[18] More important in our case, the dipolar interactions are expected to be different within the three parts, leading to a different slanting of the loops, which is the dominant feature in such dense arrays.[33] Thus, our modulated wires may be expected to reverse their magnetization in steps for individual segments, and to exhibit a stable domain boundary in a range of external fields bounded by the two distinct values of switching fields for each diameter.

**Figure 7** presents magnetic hysteresis loops recorded on a sample consisting of an ordered array of modulated NiCo nanowires embedded in their alumina matrix. The lengths of the thick, narrow and thick segments are 10 μm, 13 μm, and 10 μm, respectively. The comparison of curves recorded with the wires oriented parallel and perpendicular to the applied field highlights the anisotropy of the sample. More interestingly, the loop exhibits kinks in the parallel configuration, which define a segment of curve with large slope (near zero field) distinct from the rest of the reversal, happening with a lower slope over a broad field window between approximately –40 kA/m and +40 kA/m.

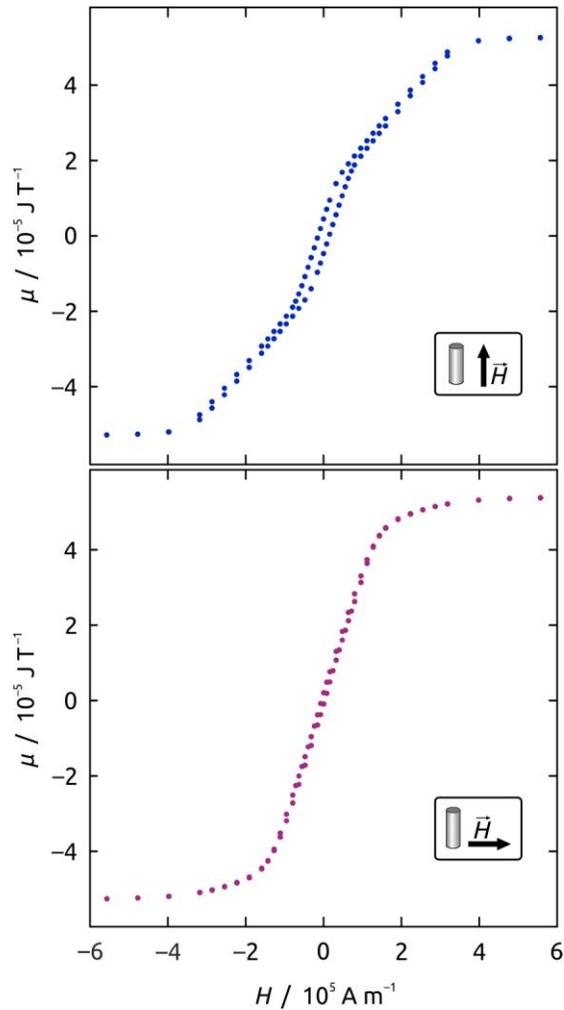

**Figure 7.** Magnetic (SQUID) hysteresis loops recorded on a sample of NiCo wires with diameters 400 nm (10 μm) / 150 nm (13 μm) /400 nm (10 μm) at room temperature with field applied parallel to the wires' long axis (top) perpendicular to it (bottom). The three parts with different slopes in the former curve are attributed to the distinct switching behavior of segments of different diameters.

With this in mind, we may attribute the magnetization change occurring around zero field to the magnetic reversal of the central, narrow segments, which are subject to a lower internal field, whereas the thicker extremities reverse more gradually. This interpretation is confirmed experimentally when three sample are compared which differ from each other solely by the length of the central segment (**Figure 8**). When this segment is varied from 3 µm to 9 µm and 13 µm, the steep section of hysteresis becomes more and more prevalent. This corresponds to the increase in the total magnetic moment of this segment associated with the increase in its length.

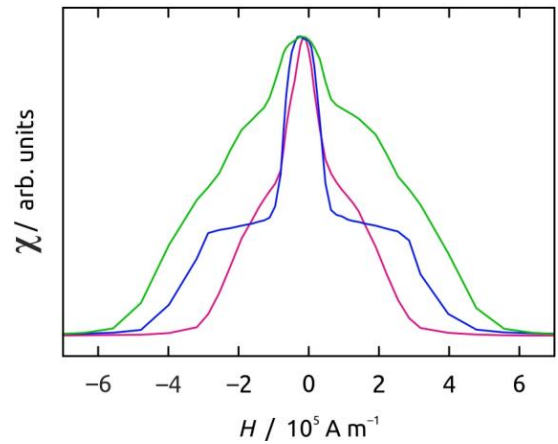

**Figure 8.** SQUID magnetic hysteresis loops presented as first derivatives (susceptibility) of three samples of modulated NiCo wires, with the field applied parallel to their long axis. The red, green and blue curves correspond to central (thin) segment lengths of 3 µm (green), 9 µm (red) and 13 µm (blue), respectively.

The distinct switching fields of the three segments can be utilized to trap magnetic domain walls at or near diameter modulations. **Figure 9** displays the morphology and magnetic stray fields of an isolated nanowire, recorded by atomic force and magnetic force microscopies, respectively. The wire was prepared by an oscillatory quasistatic demagnetization procedure performed from 1 T, with magnetic field applied out-of-plane (oop), i.e. perpendicular to the wire axis. The MFM micrograph shows the presence of contrast (stray field) at the modulations, as expected because the change of wire area implies that the excess induction gives rise to stray field. More interestingly, however, it displays a feature located near one modulation, which we can confidently attribute to a domain wall given that the two wire extremities have the same contrast (possible only if the wire consists of two domains of opposite direction). In most wires we evidence one or more such domain walls. This shows the effectiveness of the trimodulated wires to confine domain walls in a well-defined area, suitable for further investigations.

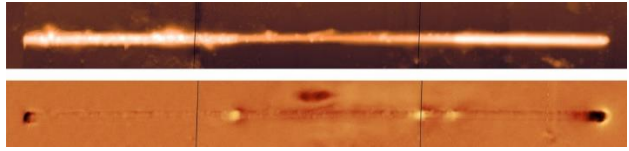

**Figure 9.** Demonstration of a magnetic domain wall pinned near a modulation. A wire isolated from the matrix displays three 10-µm long segments of distinct diameters (40 / 30 / 40 nm) in atomic force microscopy (AFM, top). The corresponding magnetic force microscopy (MFM) image (lower panel) shows the presence of magnetic stray fields at the wire extremities and diameter modulations, as expected, but also at a point situated approximately 1.5 µm from one modulation. That this stray field is caused by the presence of a domain boundary is proven by the head-to-head magnetization expressed by the other contrast points.

## 4. Modeling magnetic properties of modulated wires

### a. Arrays of wires

A global hysteresis loop is the basic characterization for a magnetic material. Even though one may in the end use local properties of single objects, a global hysteresis loop may inform about intrinsic properties such as material magnetization and sample geometry, having an impact on dipolar interactions. For instance, for a soft magnetic material, magnetization may be derived if the geometry of the system is known, through demagnetizing coefficients. In practice, this is more reliably achieved along hard-axis loops, which do not depend on hysteretic effects. **Figure 10** shows hysteresis loops of macroscopic arrays of trisegmented wires with two geometries: the central segment of diameter 120 nm and length either 3 µm or 13 µm, embedded between two segments of diameter 420 nm and length 6 µm each, all this with pitch 490 nm. We name in-plane (ip) the situation where the magnetic field is applied parallel to the membrane (i.e. across the axis of the wires), and out-of-plane (oop) the situation where the magnetic field is applied perpendicular to the membrane (i.e. along the axis of the wires). The pinpoint dipole approximation can obviously not be used to describe long wires in a dense array, as proposed in the early days of nanomagnetism.[34] Yet, there exist simple models to fit hysteresis loops in the case of straight nanowires considered uniformly magnetized, as first proposed by Wang *et al*[33], and later refined.[19,20] In short, the ip loops may be fitted with high confidence as no hysteretic event is involved, so that the analytical and exact coherent rotation model is perfectly relevant. From the fitting of the loop, $M_s$ may be extracted if $p$ is known, or the reverse. Using as fixed input the geometric parameters determined accurately by SEM, the resulting magnetization from the best fit is 862 kA/m (1.08 T in induction units). This is in perfect agreement with CoNi alloys of the present composition, CoNi being suitable for a linear interpolation on the Slater-Pauling curve.[35]

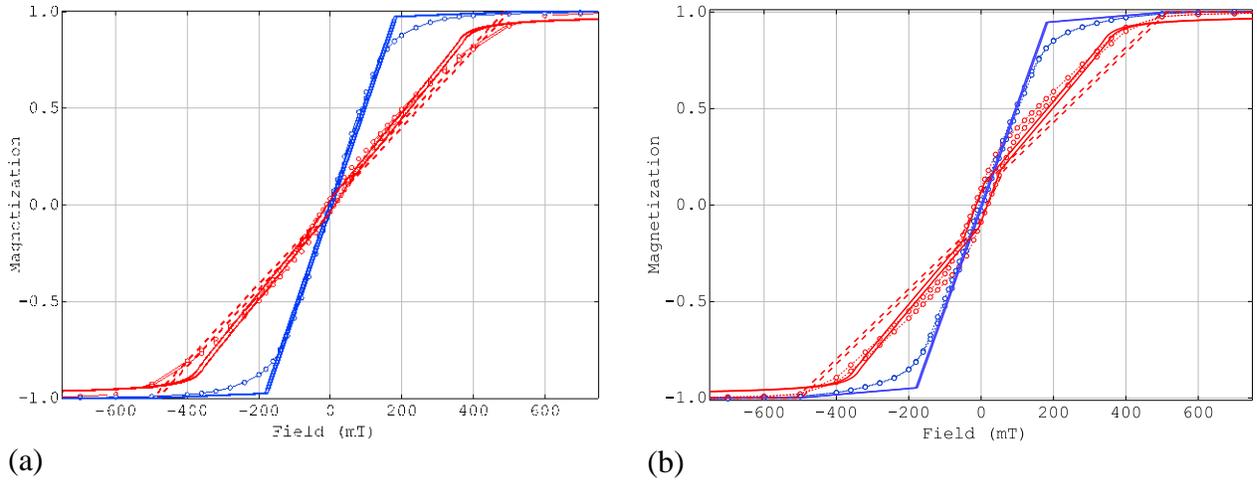

(a)             (b)

**Figure 10.** Hysteresis loops of samples with thick / narrow / thick segments of (a) 6 μm / 3 μm / 6 μm and (b) 6 μm / 13 μm / 6 μm, recorded along the in-plane (blue) and out-of-plane (red) directions. Experiments are shown with open symbols. The dotted line stands for the simple model based on uniformly magnetized wires, while the continuous line represents the curling model. The modelling parameters are: magnetization 862 kA/m; pitch 490 nm; narrow segment with diameter 120 nm and switching field 20 mT; broad segment with diameter 410 mT and switching field 5 mT.

The case of oop loops is more complex to tackle, and cannot be done so exactly, as hysteresis is involved. The wires are assumed to switch through nucleation at a wire end, and fast propagation of a domain wall. Each wire is ascribed the same individual switching field $H_c$, however the global loop is slanted as inter-wire interactions induce a dipolar contribution to the magnetic field felt by each wire. The dipolar field can be computed by considering the top and bottom charges on both sides of the membrane.[33] The saturation field is $H_\text{c} + H_\text{p}$. $H_\text{p} = \gamma p M_\text{s}$, with $\gamma \gtrsim \frac{1}{2}$ is a phenomenological parameters comprised between 0.5 and 1, taking into account nucleation effects (see Annex). For the ip direction, coherent and reversible rotation of magnetization occurs, coercivity and remanence are zero, and saturation field is $M_\text{s}\,[(1-p)/2]$.

Two complications arise here. First, we need to consider multiple segments. Second, the larger diameter of the outer segments, which gives rise to large intra-wire dipolar fields, threatens the validity of the hypothesis of uniform magnetization, and may possibly be responsible for the slow saturation of the loops (**Figure 10**), as previously reported especially for rather short wires.[36] No analytical model has been proposed to describe either of these two aspects. We will now address them in two steps and obtain a model describing such complex arrays with a remarkable accuracy.

Let us first discuss a way to deal with the diameter modulations. In our case, the outer segments have a larger diameter, thus we expect nucleation to occur at rather low field.[37] Second, the modulations are sharp and very significant in diameter, so that we expect pinning of domain walls before propagation into the central low-diameter segment.[38] Thus, in practice it is as if the central segment switches on its own under the effect of external plus all internal magnetostatic fields, albeit with a switching field smaller than an isolated wire due to the facilitated nucleation through injection of a domain wall into the wire (**Figure 11a**). Thus, the behavior of the array of trimodulated wires should be similar to the one of three independent arrays described by the model mentioned above for straight wires, each characterized by a distinct demagnetizing field. Indeed, a uniformly magnetized slab with infinite lateral dimensions (one of the three arrays, considered in the mean field as a medium with magnetization $pM_\text{s}$) gives rise to no stray field outside the slab, whatever the direction of magnetization (**Figure 11b**). So, it is as if this slab was independent. The resulting equations for hysteresis loops are provided in Annex. The dotted lines in **Figure 10** show the outcome of this first model compared with the present experimental data. The geometric parameters (length and diameter of the various segments) are fixed in the model, as determined with scanning electron microscopy. Magnetization is also fixed to the value 862 kA/m determined from the ip loops. Only the two coercive fields are let free to best adjust the experiments. The general shape of the loop is reproduced, however the match is not perfect: the slope is slightly lower in the model than in the experiments. This residual imperfection will be lifted by the second part of the model, described below.

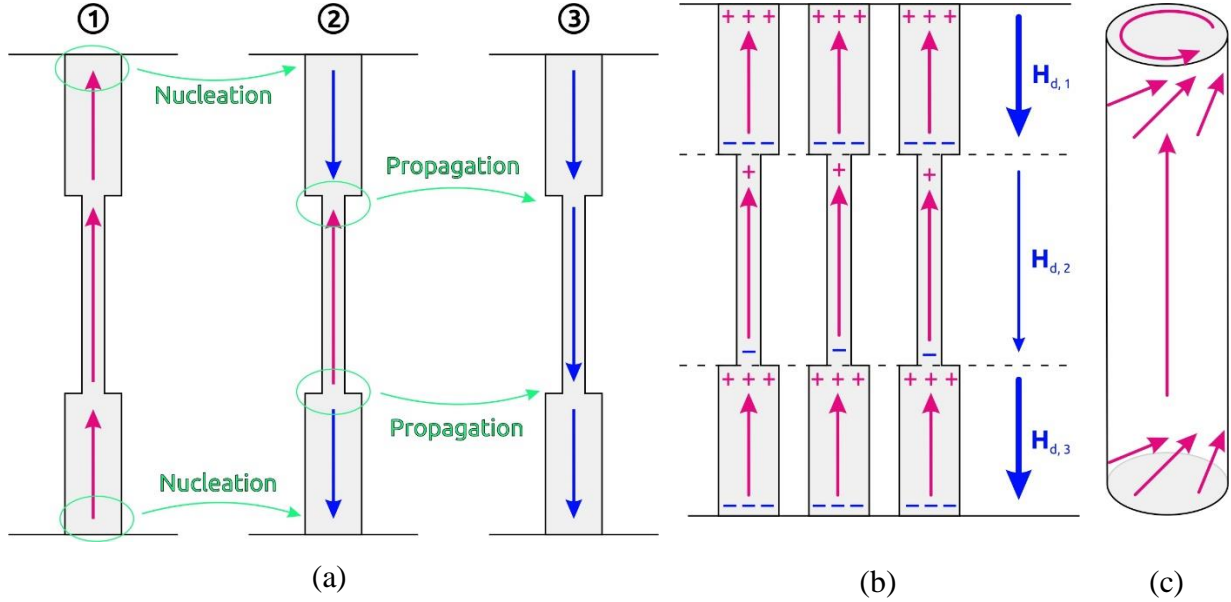

**Figure 11.** Schematic illustration of processes implemented in the modeling. (a) Nucleation at the end of the thick segments, followed by injection of a domain wall into the narrow segments. (b) Magnetic charges arising from each layer, responsible for a demagnetizing field in this sole layer. (c) Curling domain ends in thick segments.

In this second stage, we consider deviations from uniform magnetization. Non-uniform distributions of magnetization arise at wire ends, since they spread the magnetic charges and thus decrease magnetostatic energy. Above a diameter of typically seven times the dipolar exchange length (40 to 50 nm in our case), these deviations take the form of curling around the wire axis, in the case of a perfect geometry.[39,40] These so-called domain ends may be viewed as half a domain wall (**Figure 11**). In a previous publication,[41] we proposed an analytical model to predict the width of domain walls in cylindrical nanowires, based on the balance of exchange and dipolar energies. Although we estimated these energies in a handwaving fashion, the model was successful, e.g. reproducing the variation of wall width with the square of the wire diameter beyond typically 100 nm found via numerical simulation. Considering domain ends as half a domain wall, here we extend this model to include the Zeeman energy from the magnetic field, be it external or dipolar (see Annex). This field tends to compress the domain end when it is parallel to the magnetization in the body of the wire, and to stretch it when antiparallel. Thus, the curling model provides three features. First, it rounds off the hysteresis loop edges because magnetization in the domain ends is no longer strictly parallel to the applied field. Second, it brings an extra susceptibility contribution to hysteresis loops, through the compression/stretching of the domain ends. Third, it predicts the nucleation field, defined for the divergence of the width of the domain end for a given value of the applied field.

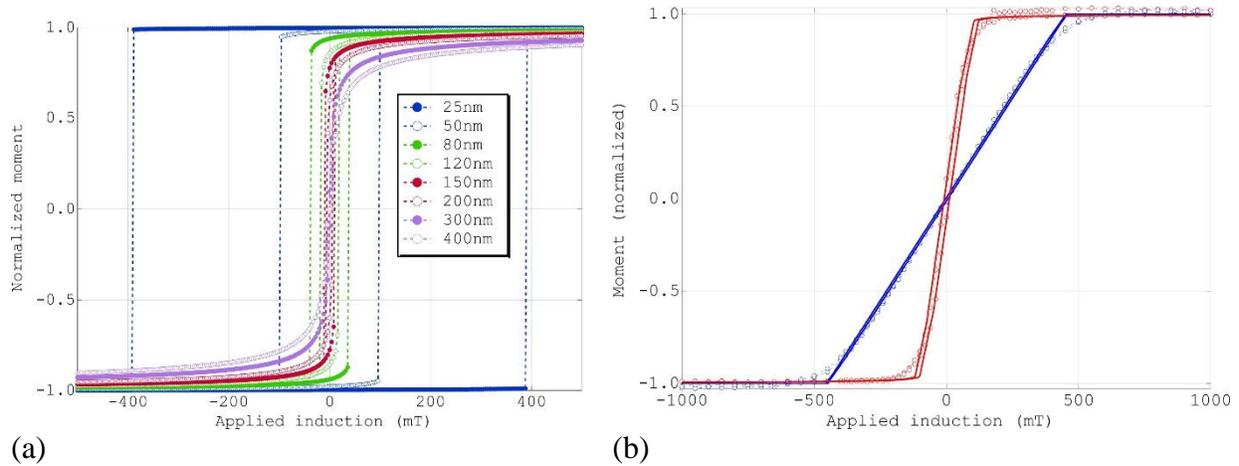

**Figure 12. (a)** Hysteresis loops predicted by the analytical curling model for various diameters, with no adjustable parameter. The fixed parameters are: magnetization 862 kA/m, exchange stiffness $1.5 \cdot 10^{-11}$ J/m, wire length 1 μm. **(b)** Hysteresis loops of trisegmented wires with diameters 200 nm / 150 nm / 200 nm, segment length 10 μm each, and pitch 410 nm, both ip (blue) and oop (red). Symbols show experiments, fitted with a line by the curling model with magnetization as the sole adjustable parameter, yielding 822 kA/m.

Thus, the model predicts the hysteresis loop of any wire with given length and diameter, both in rounding and switching value, in the absence of any adjustable parameter (**Figure 12a**). Note that curling domain ends should have no impact on ip loops, as the saturation field for a single curling wire or uniformly magnetized wire is $M_s/2$, in both cases. Curves for the refined model are shown as continuous lines on **Figure 11**. The agreement with experiments is improved, despite having now no adjustable parameter. Note the rounding at large field and the increased susceptibility at low field, resulting from the compression and the stretching of domain ends, respectively. Thus, while more refined techniques can be applied such as first-order reversal curves (FORC) and the Preisach model,[42] our model shows that the accurate analysis of a single hysteresis loop remains important to provide a quick feedback on material parameters. **Figure 12b** shows hysteresis loops for a sample with very different geometry, namely tri-segmented with diameters 200 nm, 150 nm, and 200 nm, and segment length 10 μm for all three parts. The model perfectly fits these loops with again no adjustable parameter except for a refinement of magnetization. This demonstrates that it is a robust model, suitable for very different geometries.

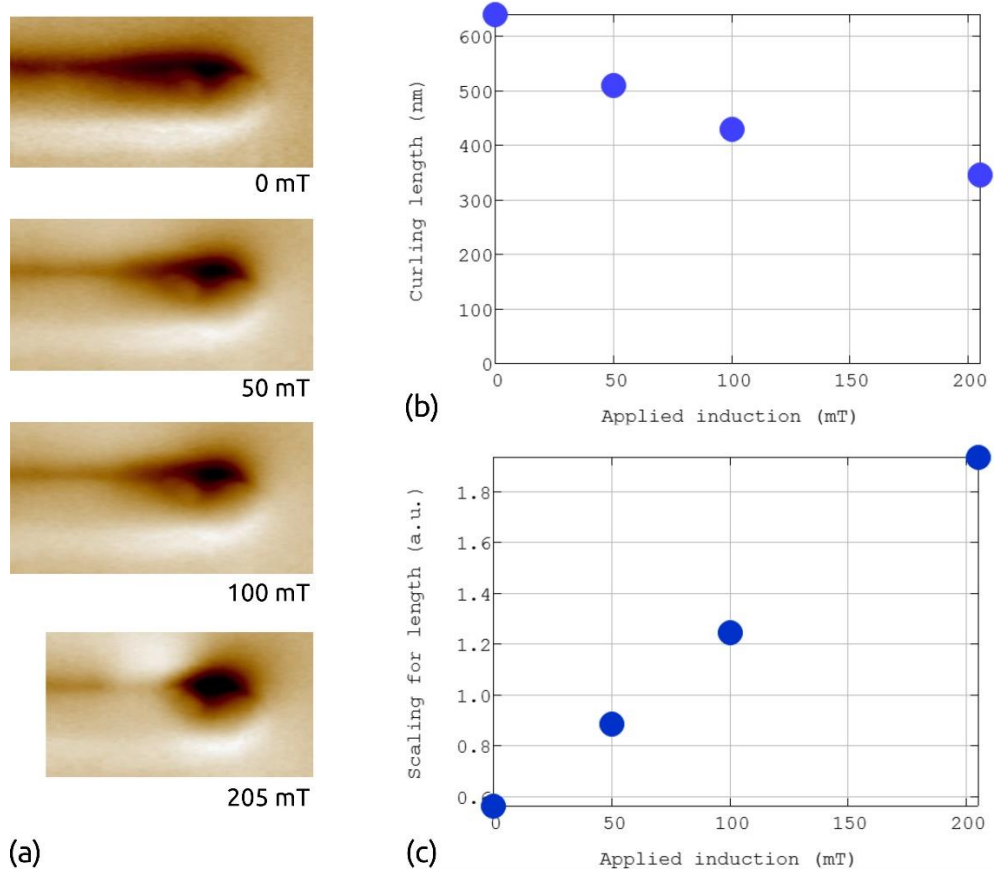

**Figure 13.** Magnetic force microscopy of a single $Co_{40}Ni_{60}$ wire with diameter 200 nm, freed from the membrane and supported on a Si surface. (a) Domain ends versus longitudinal applied field. The height of all images is 1000 nm. The opposite contrast on either side of the wire arises from the tilt of the tip axis and oscillation direction, with respect to the normal to the supporting surface[43], (b) Length of end domains as determined from the MFM images as full-width at half-maximum. (c) Scaling of experimental data shown in b, following Eq.(3).

### b. Single wires

Here, we report on the magnetic microscopy of single wires, to support experimentally the claims of the curling model introduced previously. We imaged the length $L$ of curling domain ends, identified from the distribution of charges along the wire axis, as a function of the longitudinal applied field favoring magnetization in the wire (**Figure 13a-b**). As expected, $L$ decreases with increasing applied field. Plotting the left-hand side $1/L^2$ scaling law of Eq.(3), shows a linear variation with the applied field (**Figure 13c**), although the intercept with the y axis is slightly below 1. This supports the validity of the curling model.

## Conclusions

The results presented in this paper establish a preparative strategy for the generation of smooth, cylindrical magnetic nanowires featuring well-defined modulations of diameter along their length, with for instance two segments of broad diameter encompassing one segment with narrower diameter. This design allows magnetic domain walls to be kept in the system upon quasistatic demagnetization with an oscillatory magnetic field applied across the axis of the wire. This provides a model system for studying domain walls, such as their inner structure.[44] Suitably designed modulations should provide energy barriers confining the domain wall in the central segment, which should prove useful for the investigation of domain wall mobility under magnetic field or spin-polarized conduction electrons current, without annihilating the domain wall at the end of the wire. Besides, we provide a simple yet robust analytical model fitting the in-plane and out-of-plane hysteresis loops of modulated arrays, taking into account demagnetizing field and curling effects and with no adjustable parameters. The model can be used reliably to extract geometric and/or material parameters, such as magnetization.

ANNEX

We gather here technical aspects related to the model for hysteresis loops, and the means for its implementation.

### a. Modelling hysteresis loops for arrays of straight wires

Along the oop direction, the saturation field related to inter-wire dipolar interactions is the oop demagnetizing field for a saturated sample: $pM_s$ with $p = \frac{\pi}{2\sqrt{3}}\left(\frac{d}{D}\right)^2$ the filling factor of the membrane, which is easily derived from considering the magnetic charges on both sides of the membrane, in a mean field approach.[33] Yet, the maximum interaction field relevant for hysteresis loops can be modeled as $H_p = \gamma p M_s$, with $\gamma \gtrsim \frac{1}{2}$ a phenomenological parameter taking into account nucleation effects, where mostly the charges from the membrane side opposite to that where nucleation occurs are relevant, due to a projection effect of the charges on the same side.[20] The in-plane direction is characterized by the gradual rotation of magnetization from wire axis to ip. The resulting loop is derived from the balance between the intra-wire demagnetizing field $-(M_s/2)\sin\theta$, and the inter-wire field $p(M_s/2)\sin\theta$, considering the Lorentz cavity around each wire, with $\theta$ the angle between magnetization and wire axis. Thus, the ip saturation field of the loop is $M_s\,[(1-p)/2]$.

### b. Modelling hysteresis loops with curling

Here we extend an earlier model,[41] developed to derive a scaling law for the domain wall width in cylindrical nanowires, to the present case of end domains in the presence of an axial external field. In short, we consider a progressive curling state over a distance $L$ (**Figure 11c**), and evaluate in a quite crude fashion exchange, Zeeman and magnetostatic energy. Although the validity of the estimation of the latter term is rather inaccurate for end domains much longer than the wire diameter, this approximation allows a first insight into the phenomenon at play. One can obtain a total energy such as:

$$E_t \approx 10LA + \frac{\epsilon}{3}Ld^2\mu_0 M_s H + k^2 K_d \frac{d^4}{L}\ln\left(\frac{L}{d}\right) \qquad \text{Eq(1)}$$

The first term is exchange energy, the second one is Zeeman energy, and the third one is dipolar energy. $k$ is a phenomenological scaling constant introduced to compensate the crude approximation for the dipolar energy, whose value is determined by fitting the domain wall width determined by simulation: $k \approx 0.18$. The length of the curling domain end is found by minimizing the above against $L$. Thanks to the slow variation of the logarithm function, one has:

$$L \approx \frac{k}{\sqrt{10}} \frac{d^2}{\Delta_d} \frac{1}{\sqrt{1+\frac{\epsilon}{30}\left(\frac{d}{\Delta_Z}\right)^2}} \qquad \text{Eq(2)}$$

$\Delta_d = \sqrt{2A/(\mu_0 M_s^2)}$ is the dipolar exchange length, and $\Delta_Z = \sqrt{A/(\mu_0 M_s H)}$ is the field-dependent Zeeman exchange length. The above may also be written in the form of a scaling law versus a linear variation of applied field:

$$\frac{k^2 d^4}{10 L^2 \Delta_d^2} \approx 1 + \frac{\epsilon \mu_0 M_s H d^2}{30 A} \qquad \text{Eq(3)}$$

The linearity of this scaling law is checked from the value of $L$ determined experimentally.

### c. Implementation

We use slanted functions to adjust the hysteresis loops, reflecting interaction fields. We define $m_{sl}(H, H_0)$ as the symmetric slanted non-hysteretic function saturating at $H_0$, i.e., a linear function. For an array of single segments, considered along the oop direction in a mean-field approach, the loop with rising field reads $m_{sl}(H - H_c, \gamma p M_s)$, while the one for decreasing field reads $m_{sl}(H + H_c, \gamma p M_s)$. These functions, combined with the proper interaction fields discussed in the main text, are implemented in Wavemetrics IGOR procedures.

## Acknowledgements

The research leading to these results has received funding from the European Community's Seventh Framework Program under Grant No. 309589 (M3d).